\documentclass[pra,preprint,preprintnumbers,amsmath,amssymb,superscriptaddress]{revtex4}

\usepackage[pdftex]{graphicx, color}
\usepackage{amsmath}
\usepackage{amssymb}
\usepackage{amsthm}
\usepackage{amsfonts}						%
\usepackage{graphicx}
\usepackage[english]{babel}             	
\usepackage{epstopdf}
\usepackage{graphicx}

\newcommand{\abs}[1]{\left|#1\right|}

\renewcommand{\exp}[1]{e^{#1}}
\newcommand{\avg}[1]{\left\langle #1 \right\rangle}

\newcommand{\mat}[4]{\left[\begin{array}{cc} #1 & #2 \\ #3 &  #4  \end{array}\right]}

\begin{document}

\title{Programming balanced optical beam splitters in white paint}

\date{\today}

\begin{abstract} Wavefront shaping allows for ultimate control of light propagation in multiple-scattering media by adaptive manipulation of incident waves. We shine two separate wavefront-shaped beams on a layer of dry white paint to create two enhanced output speckle spots of equal intensity. We experimentally confirm by interference measurements that the output speckle spots are almost correlated like the two outputs of an ideal balanced beam splitter. The observed deviations from the phase behavior of an ideal beam splitter are analyzed with a transmission matrix model. Our experiments demonstrate that wavefront shaping in multiple-scattering media can be used to approximate the functionality of linear optical devices with multiple inputs and outputs.\end{abstract}

\author{S. R. Huisman}
\affiliation{MESA$^+$ Institute for Nanotechnology,
University of Twente, P.O. Box 217,
7500 AE Enschede, The Netherlands. \\
$^{\star}$Present address: Institute for Molecules and Materials, Radboud University Nijmegen, Heyendaalseweg 135, 6525 AJ Nijmegen, The Netherlands.}
\author{T. J. Huisman$^{\star}$}
\affiliation{MESA$^+$ Institute for Nanotechnology,
University of Twente, P.O. Box 217,
7500 AE Enschede, The Netherlands. \\
$^{\star}$Present address: Institute for Molecules and Materials, Radboud University Nijmegen, Heyendaalseweg 135, 6525 AJ Nijmegen, The Netherlands.}
\author{S. A. Goorden}
\affiliation{MESA$^+$ Institute for Nanotechnology,
University of Twente, P.O. Box 217,
7500 AE Enschede, The Netherlands. \\
$^{\star}$Present address: Institute for Molecules and Materials, Radboud University Nijmegen, Heyendaalseweg 135, 6525 AJ Nijmegen, The Netherlands.}
\author{A. P. Mosk}
\affiliation{MESA$^+$ Institute for Nanotechnology,
University of Twente, P.O. Box 217,
7500 AE Enschede, The Netherlands. \\
$^{\star}$Present address: Institute for Molecules and Materials, Radboud University Nijmegen, Heyendaalseweg 135, 6525 AJ Nijmegen, The Netherlands.}
\author{P. W. H. Pinkse}\email{p.w.h.pinkse@utwente.nl, www.adaptivequantumoptics.com}
\affiliation{MESA$^+$ Institute for Nanotechnology,
University of Twente, P.O. Box 217,
7500 AE Enschede, The Netherlands. \\
$^{\star}$Present address: Institute for Molecules and Materials, Radboud University Nijmegen, Heyendaalseweg 135, 6525 AJ Nijmegen, The Netherlands.}

\email{p.w.h.pinkse@utwente.nl} 
\homepage{www.adaptivequantumoptics.com}


\maketitle


\section{Introduction}
Linear optical components like lenses, mirrors, polarizers and wave plates are the essential building blocks of optical experiments and applications \cite{Hecht2002, Saleh2007}. One has often little freedom to modify the linear optical circuit, besides rearranging components or including adaptive elements. For computational applications and on-chip light processing \cite{Nielsen2000, Knill2001, Kok2007, OBrien2009}, it would be fantastic to have a programmable linear optical circuit that can be controlled during operation. Spatial light modulation in combination with random light scattering provides an excellent platform to accomplish this \cite{Freund1990}. The collective interference of scattered light propagating through a scattering material results into a speckle pattern. In wavefront shaping \cite{Vellekoop2007, Mosk2012} incident light is modulated by a spatial light modulator to obtain a desired speckle pattern for functionality. Although one has to tolerate some losses, this technique makes it possible to transform opaque scattering media in linear optical elements that are flexible in performance \cite{Vellekoop2010, Putten2011, Aulbach2011, Katz2011, McCabe2011, Guan2012}. If one controls light propagation inside strongly scattering media, one can make the most exotic linear optical circuits within a fraction of a mm$^3$.  

In earlier wavefront-shaping experiments, multiple target speckle spots have been simultaneously optimized with a single incident wavefront \cite{Vellekoop2007, Guan2012, Popoff2010}. These experiments essentially demonstrate $1 \times m$ linear optical circuits with $1$ incident mode projected to $m$ output modes. If one is capable to manipulate $n$ incident modes with wavefront shaping, it becomes possible to program $n \times m$ optical circuits with a desired transmission matrix $\textbf{T}$. To our knowledge, no experiment demonstrating this capability has been reported.

In this article we describe an experiment in which a layer of white paint is used as an optical beam splitter. We apply wavefront shaping to intensity enhance 2 output speckle spots for 2 separate incident modes, forming a $2 \times 2$ optical circuit. The enhanced speckle spots are almost correlated like the outputs of a 50:50 beam splitter. This behavior is verified by an optical interference experiment. Our measurements indicate that the system approaches the behavior of an ideal beam splitter when the intensity enhancement $\eta$ increases, which is confirmed by simulations. This surprisingly suggests that one can program beam splitters without prior measurement of the transmission matrix \textbf{T}. We explain this with random matrix calculations. Our experiment demonstrates that wavefront shaping can be extended to multiple incident modes that can interfere in a controlled manner, allowing for programmable linear optical circuits. 
\section{Interference on a beam splitter}
\begin{figure}[t!]
  \includegraphics[width=12.5 cm]{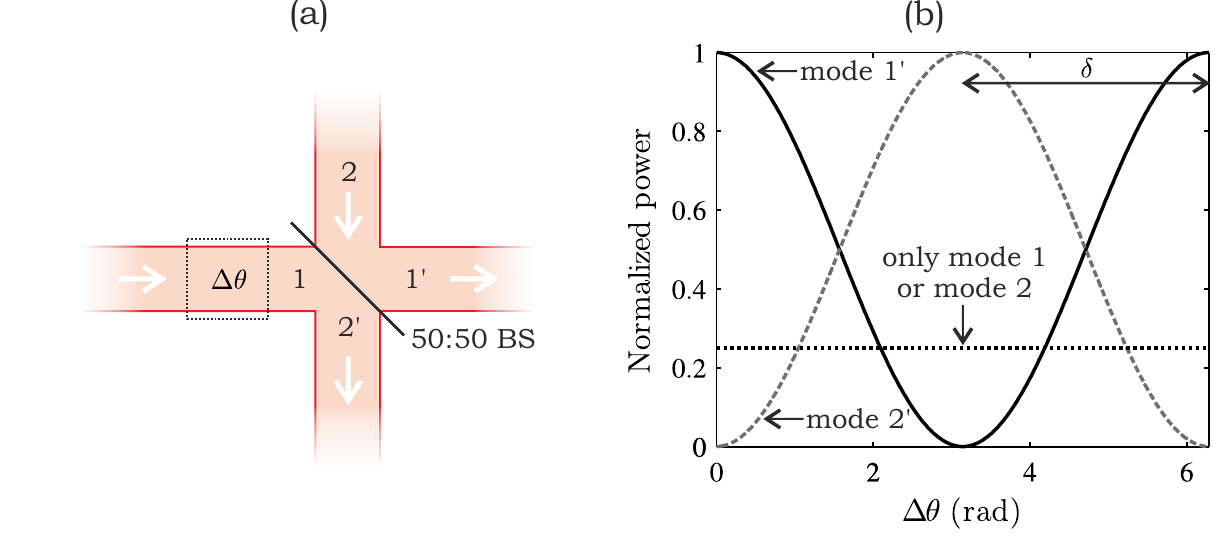}
\caption{(color online) \textbf{Interference at an ideal lossless 50:50 beam splitter.} $(a)$ Two incident modes $(1,2)$ with equal intensity interfere at a 50:50 beam splitter (BS) resulting in two output modes $(1^{\prime},2^{\prime})$. $(b)$ Normalized power in the output modes as a function of relative phase shift $\Delta \theta$ applied to incident mode $1$. All powers are normalized to the sum of the two input powers $P_1+P_2$.}
\label{Fig_7_2}
\end{figure}
The scattering matrix of a lossless beam splitter represents a unitary transformation that can be written in its most general form as the product of three matrices \cite{Campos1989}:
\begin{equation}
\bold{S}= \mat{e^{i \frac{\Psi}{2}}}{0}{0}{e^{-i \frac{\Psi}{2}}} \mat{\cos \frac{\Theta}{2}}{\sin \frac{\Theta}{2}}{-\sin \frac{\Theta}{2}}{\cos \frac{\Theta}{2}} \mat{e^{i \frac{\Phi}{2}}}{0}{0}{e^{-i \frac{\Phi}{2}}}
\label{Eq_7_1}.
\end{equation}
The nonzero terms in the first and last matrix are phase differences $\Psi$ and $\Phi$ applied by the beam splitter on the incident and outgoing modes respectively. In this article we use the term input mode for an incident wave that describes a single orthogonal input of the normal beam splitter or wavefront-shaped beam splitter. The phase angle $\Theta$ in the center matrix determines the splitting ratio, which has to be $\Theta =(2n+1)\pi/2$ for a 50:50 beam splitter, with \textit{n} an integer value. 

Now consider the experiment in Fig. \ref{Fig_7_2}$(a)$ in which two modes $(1,2)$, carrying coherent light of equal power, frequency, polarization, and wavefront are incident on a 50:50 beam splitter giving two output modes $(1^{\prime},2^{\prime})$. The relative phase difference $\Delta \theta$ between mode $1$ and $2$ can be controlled; this is the same as controlling $\Phi=\Delta \theta$ in Eq. (\ref{Eq_7_1}). The power $P_{1^{\prime}}, P_{2^{\prime}}$ as a function of $\Delta \theta$ is shown in Fig. \ref{Fig_7_2}$(b)$. $P_{1^{\prime}}$ and $P_{2^{\prime}}$ will oscillate as a function of $\Delta \theta$ with a phase difference of $\delta=\pi$ for the ideal lossless beam splitter, independent of phases $\Phi$ and $\Psi$ in Eq. (\ref{Eq_7_1}). For this graph we have set $\Psi=0$. Any nonzero value for $\Psi$ will provide a phase offset to both the output modes, essentially shifting both $P_{1^{\prime}}$ and $P_{2^{\prime}}$ along the horizontal axes by the same amount. If one of the incident modes is blocked, a constant power is detected in both output modes that is 4 times lower than the maximum power in one of the output modes when both inputs are present. Note that energy is conserved: $P_{1^{\prime}}+P_{2^{\prime}}=P_1+P_2$, and a fringe visibility of $100\%$ is observed. 

\begin{figure}[]
  \includegraphics[width=12.5 cm]{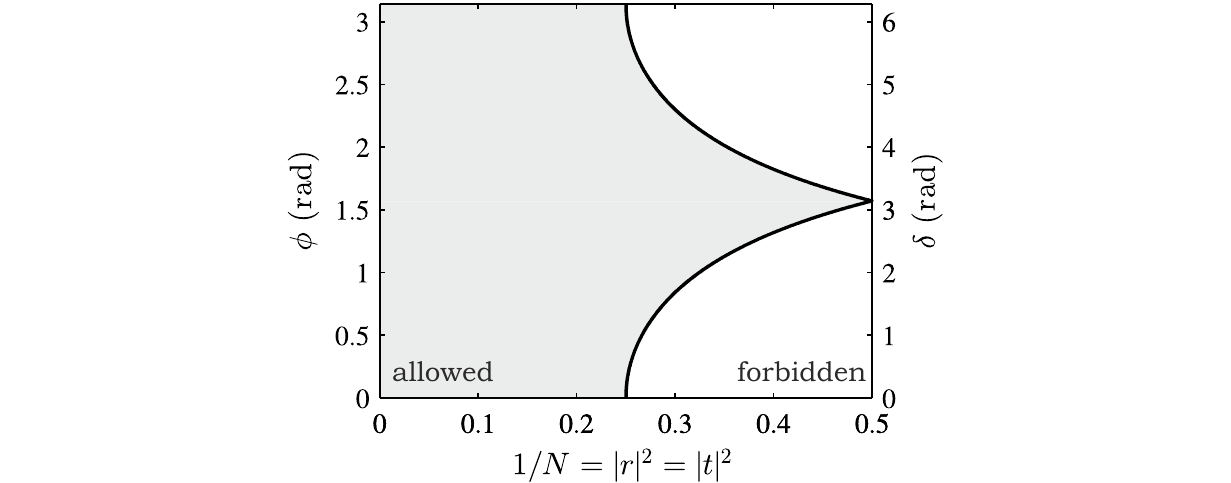}
\caption{(color online) \textbf{The allowed phase difference $\phi$ and $\delta = 2 \phi$ as a function of reflectivity $\abs{r}^2$ for the balanced beam splitter.} The white region is forbidden because of energy conservation.}
\label{Fig_7_3}
\end{figure}

Modeling lossy beam splitters is an interesting subject on its own with a wide variety of approaches \cite{Kiss1995, Leonhardt1997, Barnett1998, Knoll1999, Jeffers2000}. We describe the effective transmission matrix \textbf{T} as a non-unitary version of Eq. (\ref{Eq_7_1}). The phase difference $\delta=\pi$ will still hold if there are losses in any of the input or output modes. In such a case, the damping can be modeled in the left or right matrix in Eq. (\ref{Eq_7_1}) by making them non-unitary with a determinant smaller than 1. This is valid for the typical beam splitter one uses. However, when the scattering in the beam splitter does not conserve energy (\textit{i.e.} the damping is in the center matrix in Eq. (\ref{Eq_7_1}) dictating that not all energy goes to the considered output modes), $\delta$ could in principle take on any value as long as the total output power does not exceed the total input power. To model the most general form of a lossy beam splitter we make the following assumptions: 
\paragraph*{Assumption 1}
The system is described by an effective transmission matrix $\bold{T}$ consisting of $2\times 2$ elements.
\paragraph*{Assumption 2}
Transmission matrix $\bold{T}$ provides an equal power splitting ratio. 
\paragraph*{Assumption 3}
The power losses for both input modes are identical.
\\
\\
The system behaves as a lossy beam splitter with equal splitting ratio. For convenience to compare to the ideal beam splitter, we consider that a single input mode is reflected with power reflectance $\abs{r}^2$ and transmitted with transmission $\abs{t}^2$. For a lossy balanced beam splitter this means $\vert r \vert = \vert t \vert$ and $\abs{r}^2+\abs{t}^2\leq 1$. We relate the input amplitudes $A_1$ and $A_2$ to the output amplitudes $A_{1^{\prime}}$ and $A_{2^{\prime}}$ as $A_1\rightarrow\vert r \vert (A_{1^{\prime}}+e^{i \phi_1}A_{2^{\prime}})$ and $A_2\rightarrow\vert r \vert (e^{i \phi_2} A_{1^{\prime}}+A_{2^{\prime}})$, with phase terms $\phi_1$ and $\phi_2$. Now we set $\abs{r}^2=1/N$, with splitting factor $N$ and $N\geq 2$. This leads to the following transmission matrix:
\begin{equation}
\bold{T}= \frac{1}{\sqrt{N}}\mat{1}{e^{i \phi_1}}{e^{i \phi_2}}{1}.
\label{eq:matrix_lossy_splitter}
\end{equation}
For the ideal lossless 50:50 beam splitter, $\bold{T}$ is only unitary when $N=2$, with for example $\phi_1=\phi_2=\pi/2$, and would always result to interference as shown in Fig. \ref{Fig_7_2}$(b)$. The eigenvalues $\lambda_1$ and $\lambda_2 $ of the matrix in Eq. (\ref{eq:matrix_lossy_splitter}) are:
\begin{equation}
\lambda_1=\frac{1+e^{i\frac{\phi_1+\phi_2}{2}}}{\sqrt{N}}, \quad 
\lambda_2=\frac{1-e^{i\frac{\phi_1+\phi_2}{2}}}{\sqrt{N}}.
\label{eq:eigenvalues_lossy_splitter}
\end{equation}
Next we make the substitution $\phi=(\phi_1+\phi_2)/2$. The observed phase difference in an interference experiment is given by $\delta=\phi_1+\phi_2=2\phi$. Since $\bold{T}$ is a square matrix, from the singular values of Eq. (\ref{eq:matrix_lossy_splitter}) $\tau^2_1=\abs{\lambda_1}^2=(2+2\cos(\phi))/N$ and $\tau^2_2=\abs{\lambda_2}^2=(2-2\cos(\phi))/N$ we obtain:
\begin{equation}
\tau_1^2+\tau_2^2=\frac{4}{N}.
\label{eq:sing_value_lossy_splitter}
\end{equation}
If this relation is fulfilled, it is guaranteed that $\abs{r}^2=\abs{t}^2$. In addition $\tau_1,\tau_2\leq 1$ to guarantee a transmission not exceeding $1$. This restricts the possible $\phi$ to be in the range between $\cos^{-1}\left(N/2-1\right) \leq \phi \leq \cos^{-1}\left(1-N/2\right)$ for $2 \leq N \leq 4$, as marked by the gray area in Fig. \ref{Fig_7_3}. 

With wavefront shaping it is possible to use a multiple-scattering material as a balanced beam splitter, which is inherently lossy. In our experiment we are working with $N \gg 10^2$ and therefore any $\phi$ and $\delta$ are allowed. The scattering statistics of the sample, such as the the singular value distribution, and the intensity enhancement defining $N$ in Eq. (\ref{eq:sing_value_lossy_splitter}), determine the combination of $\tau_1$ and $\tau_2$ that satisfy Eq. (\ref{eq:sing_value_lossy_splitter}). Therefore one would not expect in general a constant probability distribution for $\delta$ in the gray marked area in Fig. \ref{Fig_7_3}. We would like to approximate the behaviour of a beam splitter where $\delta \rightarrow \pi$ since this mimics the beam splitter one normally uses.

\section{Optimization algorithm}
\begin{figure}[t!]
  \includegraphics[width=12.5 cm]{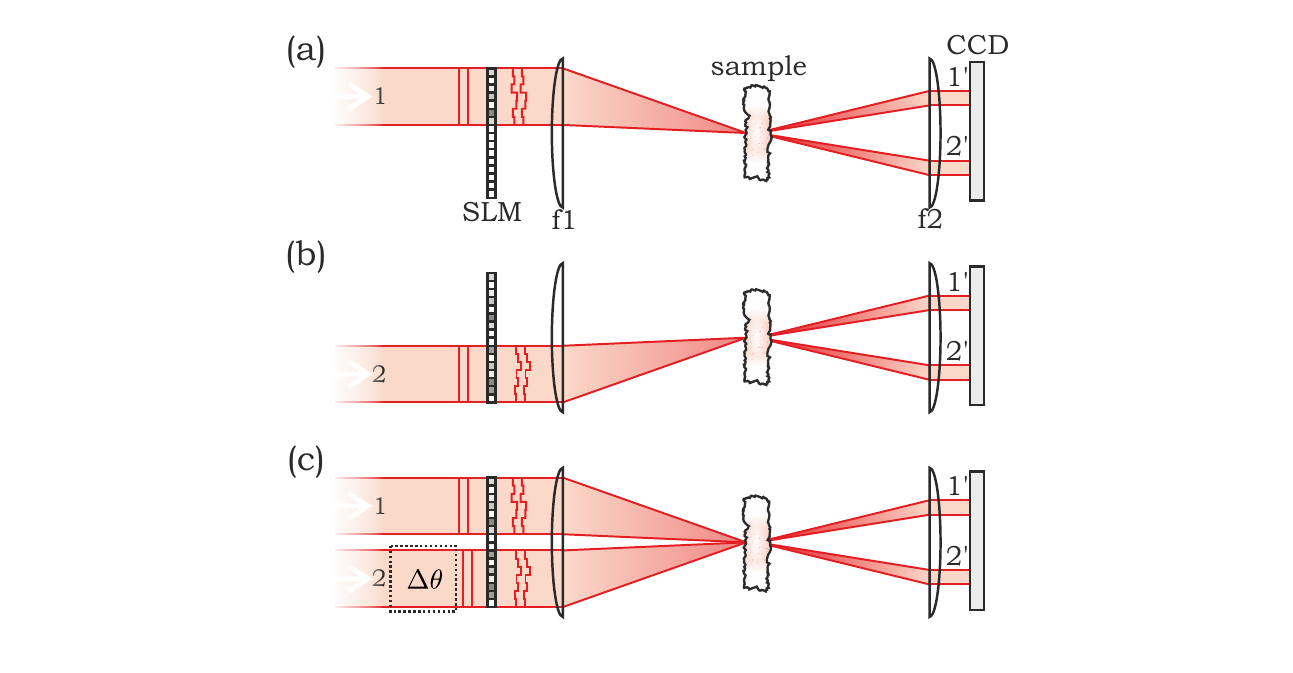}
\caption{(color online) \textbf{Optimization procedure for wavefront-shaped balanced beam splitters.} A single SLM is divided into two sections to phase modulate incident modes $1$ and $2$ that are spatially separated. $(a)$ Only mode $1$ is incident and mode $2$ is blocked. Two target speckles $1^{\prime}$ and $2^{\prime}$ are optimized on a CCD camera.
$(b)$ Only mode $2$ is incident and mode $1$ is blocked, and the same target speckles $1^{\prime}$ and $2^{\prime}$ are optimized. 
$(c)$ Both modes $1$ and $2$ are incident on the SLM, using both optimized phase patterns. A relative phase differenced $\Delta \theta$ between mode $1$ and $2$ is applied with the SLM to confirm optical interference like in Fig. \ref{Fig_7_2}$(b)$.}
\label{Fig_7_4}
\end{figure}
The optimization procedure for wavefront-shaped balanced beam splitters is illustrated in Fig. \ref{Fig_7_4}. 
We start with a single incident mode. All segments of a phase-only spatial light modulator (SLM) are subsequently addressed. The phase $\phi_n \subset \left[0, 2\pi \right)$ of the $n^{\rm{th}}$ segment is randomly chosen and the output powers $P_{1^\prime}$ and $P_{2^\prime}$ are monitored. $\phi_n$ is accepted and kept on the SLM if the summed output powers of both spots has increased and the difference power has decreased:
\begin{enumerate}
\item $P_{1^{\prime}, \rm{new}}+P_{2^{\prime}, \rm{new}}> P_{1^{\prime}, \rm{old}}+P_{2^{\prime}, \rm{old}}+\epsilon_1$ 
\item $\vert P_{1^{\prime}, \rm{new}}-P_{2^{\prime}, \rm{new}} \vert < \vert P_{1^{\prime}, \rm{old}}-P_{2^{\prime}, \rm{old}} \vert + \epsilon_2$
\end{enumerate}
with positive tolerances $\epsilon_1, \epsilon_2 \rightarrow 0$ to compensate for noise. Otherwise the previous $\phi_n$ was restored. Next the $(n+1)^{\rm{th}}$ segment is addressed, etc. After the final segment has been addressed, the entire optimization is repeated until the desired convergence is reached. 

The same procedure is repeated for the second incident mode for the same two target spots. Finally both modes are incident to perform an interference experiment as described in Fig. \ref{Fig_7_2}, where the relative phase $\Delta \theta$ is controlled with the SLM.

We have decided to select this optimization algorithm because of ease of implementation and the guaranteed convergence to spots of equal power. There are algorithms that work faster and more efficiently resulting in speckles of higher intensity enhancement \cite{Vellekoop2008-2}. At any rate, in the next sections we demonstrate that our algorithm is adequate for this experiment.

\section{Experimental setup}
\begin{figure}[t!]
  \includegraphics[width=12.5 cm]{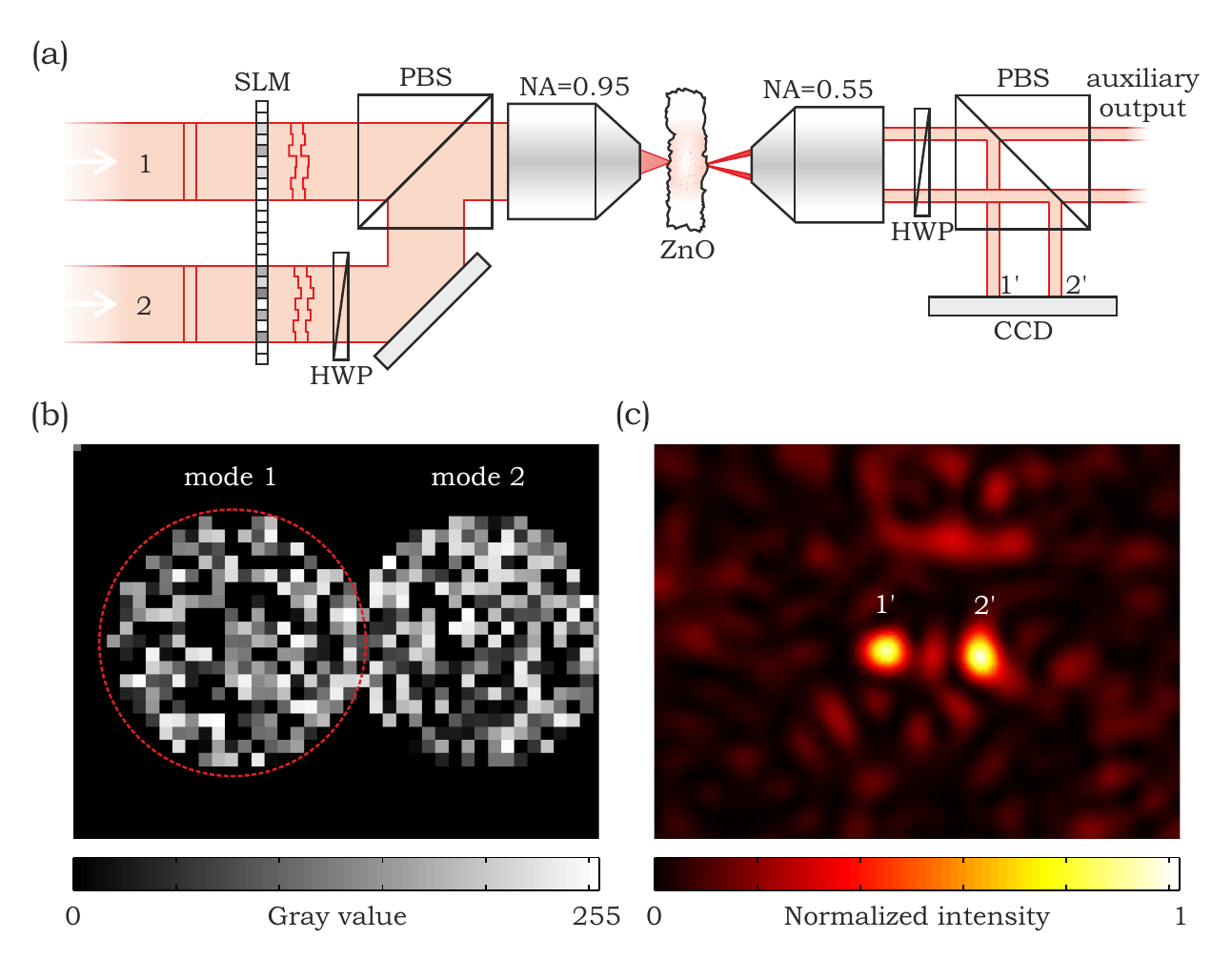}
\caption{(color online) \textbf{Setup for wavefront-shaped balanced beam splitters.} $(a)$ Two incident modes $(1,2)$ are phase-modulated with a spatial light modulator (SLM). Both modes are spatially overlapped with a polarizing beam splitter cube (PBS). The modes are focused on a layer of white paint (ZnO particles) that has been spray coated on a 1 mm thick microscope slide. The transmitted light is projected on a CCD camera. Two output modes $1^{\prime}$ and $2^{\prime}$ are selected. $(b)$ Optimized phase pattern on the SLM. $(c)$ Camera image for two optimized speckles when mode 1 is blocked.}
\label{Fig_7_5}
\end{figure}
The setup is illustrated in Fig. \ref{Fig_7_5}$(a)$. The light source is a mode-locked Ti:Sapphire laser (Spectra-Physics, Tsunami) emitting transform-limited pulses at a repetition rate of $80$ MHz with a pulse width of approximately $0.3$ ps and a center wavelength of $790.0$ nm. The pulses are spectrally filtered by a Fabry-Perot cavity with a linewidth of 1.5 nm. The beam is split and coupled into two separate single-mode fibers. The output modes have identical polarization and waist and form the input modes $1$ and $2$. The two modes are phase-modulated with a SLM (Hamamatsu, LCOS-SLM). The two modes are spatially overlapped with a half-wave plate (HWP) and polarizing beam splitter (PBS) cube, resulting in co-linear propagation of modes with orthogonal polarization. This allows us to completely fill the aperture of the objective (NA=0.95, Zeiss) that focuses the light on a layer of white paint. Both pulses arrive simultaneously at the sample to within $20$ fs. We make sure that the power of both incident modes on the objective are identical (power of approximately $0.5$ mW per mode). The layer of white paint consists of ZnO powder with a scattering mean free path of $0.7 \pm 0.2$ $\mu$m. The layer is approximately $30$ $\mu$m thick and spray painted on a glass microscope slide of 1 mm thickness. The transmitted speckle pattern is collected with a second objective (NA=0.55, Nikon) and imaged on a CCD camera after reflection on a PBS, see for example Fig. \ref{Fig_7_5}$(c)$. The intensity values for the CCD pixels that correspond to the target speckle spots are integrated to obtain the output powers for the enhanced speckle spots. The optimized speckles can be transmitted through the PBS, towards a different part of the setup for applications, by rotating the HWP. 

The SLM was divided into segments of 20x20 pixels. Each segment was sequentially addressed with a random phase as described in the optimization algorithm. The phase is applied by writing a gray value between 0 and 255. This corresponds to a phase modulation depth of $(2.0 \pm 0.1) \pi$ rad. This algorithm was repeated approximately 15 times for all segments to obtain two enhanced speckles of equal power at $1^{\prime}$ and $2^{\prime}$, see Fig. \ref{Fig_7_5}$(c)$. The total optimization procedure for both incident modes takes about $3$ hours. We confirm interference between the output modes by adding a phase offset to the encircled area in Fig \ref{Fig_7_5}$(b)$. 

\begin{figure}[]
  \includegraphics[width=12.5 cm]{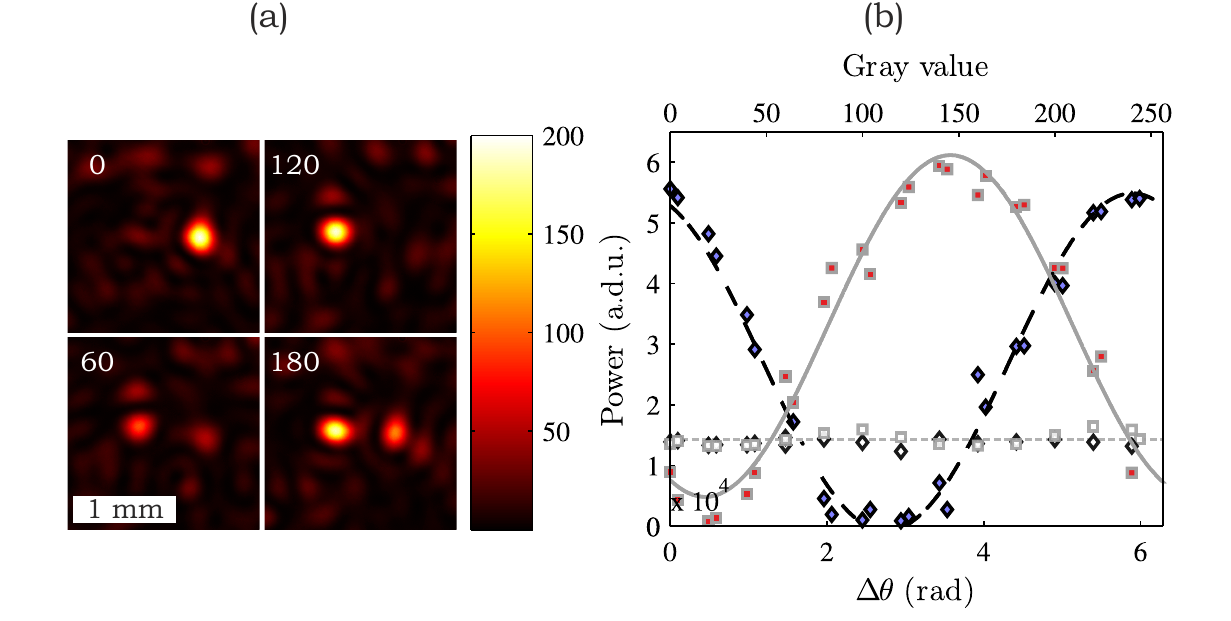}
\caption{(color online) \textbf{Interference between two optimized speckles of a wavefront-shaped beam splitter.} A phase shift is applied on one of the incident modes by applying an offset to the phase pattern of the SLM. $(a)$ Camera images for different phase offsets $\Delta \theta$ expressed as gray values. $(b)$ Power in the two optimized speckles (red squares and blue diamonds) as a function of gray value offset. The solid and dashed lines are sinusoidal fits. When only one incident mode is present, a constant power in both target speckles is observed (white diamonds and white squares).}
\label{Fig_7_6}
\end{figure}

\section{Experimental results}
Figure \ref{Fig_7_6} presents the main result of this article: optical interference with a wavefront-shaped nearly balanced beam splitter. Figure \ref{Fig_7_6}$(a)$ shows a series of camera images when on incident mode $1$ a phase offset is applied. The number in the left top corners represent the phase offset $\Delta \theta$ in gray values. All pictures are subsequently taken with the same integration time. The intensity clearly oscillates between the two target spots. 

Figure \ref{Fig_7_6}$(b)$ shows the output power $P_{1^{\prime}}$ (red squares) and $P_{2^{\prime}}$ (blue diamonds) as a function of the applied phase difference $\Delta \theta$. Both curves show sinusoidal behavior and are approximately out-of-phase, mimicking the behavior of an ideal beam splitter as is shown in Fig. \ref{Fig_7_2}$(b)$.  We expect an error in the phase of about $\Delta(\Delta \theta) = 0.1$ rad due to interferometric stability during data collection and an additional systematic error of $0.1$ rad due to phase calibration (both not shown). We have fitted two functions of the form $A \sin(\Delta \theta+b)+c$ to the measured power, which is in good agreement with the data points. From $b$ we determined the phase difference $\abs{\delta} = 2.30 \pm 0.14$ rad, close to but significantly different from the value of $\delta=\pi$ of an ideal beam splitter. Both $P_{1^{\prime}}$ and $P_{2^{\prime}}$ show a fringe visibility of approximately $100\%$, which indicates a near-perfect mode matching between the output modes for the two seperate incident modes. The maximum measured power in both spots is approximately the same to within $5\%$. When one of the incident modes is blocked in the interference experiment, the output power is approximately constant (white diamonds and squares). The small spatial separation between mode $1$ and $2$ on the SLM gives a small crosstalk, causing fluctuations within $10 \%$. The output power is approximately 4 times lower than the maximum power in one output mode when both input modes are incident, in excellent agreement with Fig. \ref{Fig_7_2}$(b)$. 

\begin{figure}[t!]
  \includegraphics[width=12.5 cm]{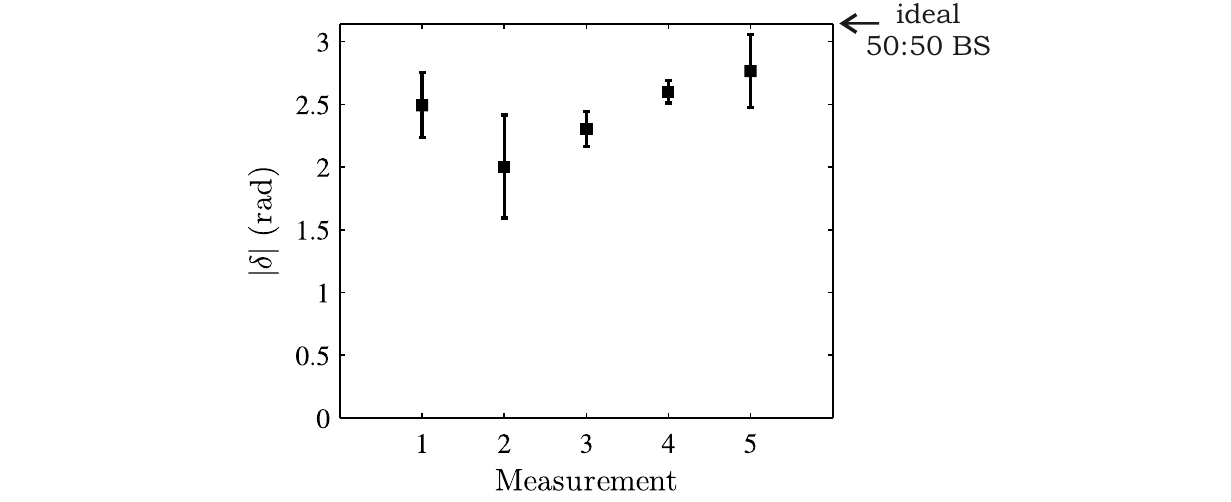}
\caption{\textbf{Measured phase difference $\abs{\delta}$ for different realizations of the wavefront-shaped balanced beam splitter.}}
\label{Fig_7_7}
\end{figure}

We have repeated this interference experiment 5 times for different target speckle spots and determined $\abs{\delta}$. The result is shown in Fig. \ref{Fig_7_7}. All measurements were performed under comparable circumstances. Although the number of measurements are not sufficient for any statistical relevant conclusion, our measurements suggest a tendency for  $\abs{\delta}$ to cluster close to $\pi$. In the next section we present a model that predicts this behavior.  

\section{Model for the phase difference}

\begin{figure}[t!]
  \includegraphics[width=12.5 cm]{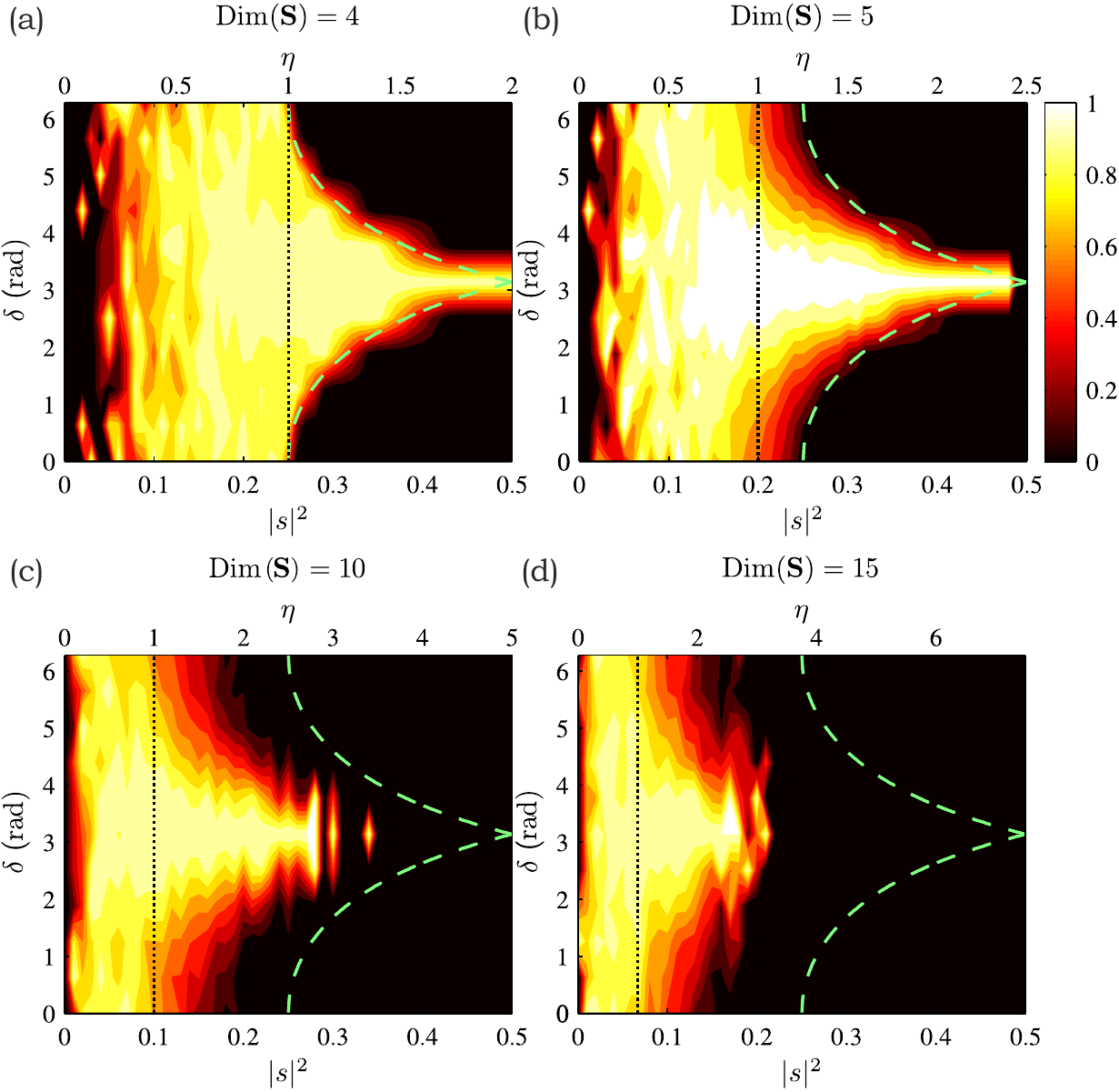}
\caption{\textbf{Calculated normalized probability distribution for the phase difference $\delta$ for random scattering matrices $\textbf{S}$ with different $\rm{Dim}(\textbf{S})$ containing balanced lossy beam splitters.} All 4 elements in the scattering matrix that form the beam splitter have an equal amplitude within $1\%$ to be accepted. The green dashed line represents the boundary of allowed and forbidden phases as in Fig. \ref{Fig_7_3}. The black vertical dashed line represents $\avg{\abs{s_{a,b}}^2}$.}
\label{Fig_7_8}
\end{figure}

In this section we model the observed phase difference $\delta$ in the interference experiment based on random matrix theory. Light undergoes  isotropic multiple scattering in a sample with a thickness much larger than the scattering mean free path $l$ and $k l \gg 1$. We therefore expect the transmission matrix $\bold{T}$ to follow the statistics of a random matrix, as was demonstrated experimentally for ZnO \cite{Popoff2010}. One could argue that $\bold{T}$ is a subset of the scattering matrix $\textbf{S}$, and therefore $\phi_1$ and $\phi_2$ can take in principle any value between $\left[ 0, 2\pi \right)$ with equal probability. This would naively result in a constant probability distribution for $\delta$, which is not observed.

The scattering matrix \textbf{S} has to be unitary, which sets restrictions on the allowed values for each element $s_{a,b}$. Consider a random \textbf{S} in a basis where one input mode is one element of the input vector and one target output speckle spot is one element of the output vector. If $\textbf{S}$ contains a beam splitter of equal splitting ratio, there have to be 2 rows and 2 columns in $\textbf{S}$ with corner elements of approximately the same amplitude: 
\begin{equation}
\abs{s_{i,j}} = c_1 \abs{s_{n,j}}= c_2 \abs{s_{i,m}}=c_3 \abs{s_{n,m}}
\end{equation}
with $\{i,j,m,n\} \leq N$ positive integers and $c_1 \approx c_2 \approx c_3$. For a random scattering matrix with dimension $\rm{Dim}(\textbf{S})=2$, the only possibility for a balanced beam splitter is that $\delta =\pi$. From matrix algebra it follows that for $\rm{Dim}(\textbf{S})=3$, $\delta$ can only lie on the boundary lines of the gray area of Fig. \ref{Fig_7_3} and the amplitude coefficients of $\textbf{S}$ should satisfy $\abs{s_{a,b}}^2 \geq 1/4$. For $\rm{Dim}(\textbf{S}) \geq 4$ any phase becomes accessible within the gray marked area of Fig. \ref{Fig_7_3}. However, the corresponding phase distribution is strongly dependent on $\rm{Dim}(\textbf{S})$, as illustrated in Fig. \ref{Fig_7_8}. There we have generated many random scattering matrices with different dimension that contain a balanced beam splitter.\footnote{We use Matlab 2013 for creating random unitary matrices $\textbf{S}$ with dimension $N$ using the following syntax:\\
Random\_Matrix=rand($N$,$N$)+1i*rand($N$,$N$); \%generates uniformly distributed random complex numbers\\
$[$\textbf{S}, r$]$=qr(Random\_Matrix); \% \textbf{S} is the output of an orthogonal-triangular decomposition.} The corresponding intensity enhancement is given by $\eta = \abs{s}^2/\avg{\abs{s_{a,b}}^2}$, with $\abs{s}^2=(1/4)(\abs{s_{i,j}}^2+\abs{s_{n,j}}^2+\abs{s_{i,m}}^2+\abs{s_{n,m}}^2)$, and $\avg{\abs{s_{a,b}}^2}=1/\rm{Dim}(\textbf{S})$. An increased probability for $\delta = \pi$ is observed with higher $\eta$. The probability distribution becomes flat for small $\eta$. How this scales depends strongly on $\rm{Dim}(\textbf{S})$. It becomes extremely difficult to observe $\eta>4$ for $\avg{\abs{s_{a,b}}^2}<0.01$, because the probability to get these realizations out of random unitary matrices becomes astronomically small. Our experiment, however, is setup to achieve exactly such realizations by optimization.

\begin{figure}[t!]
  \includegraphics[width=12.5 cm]{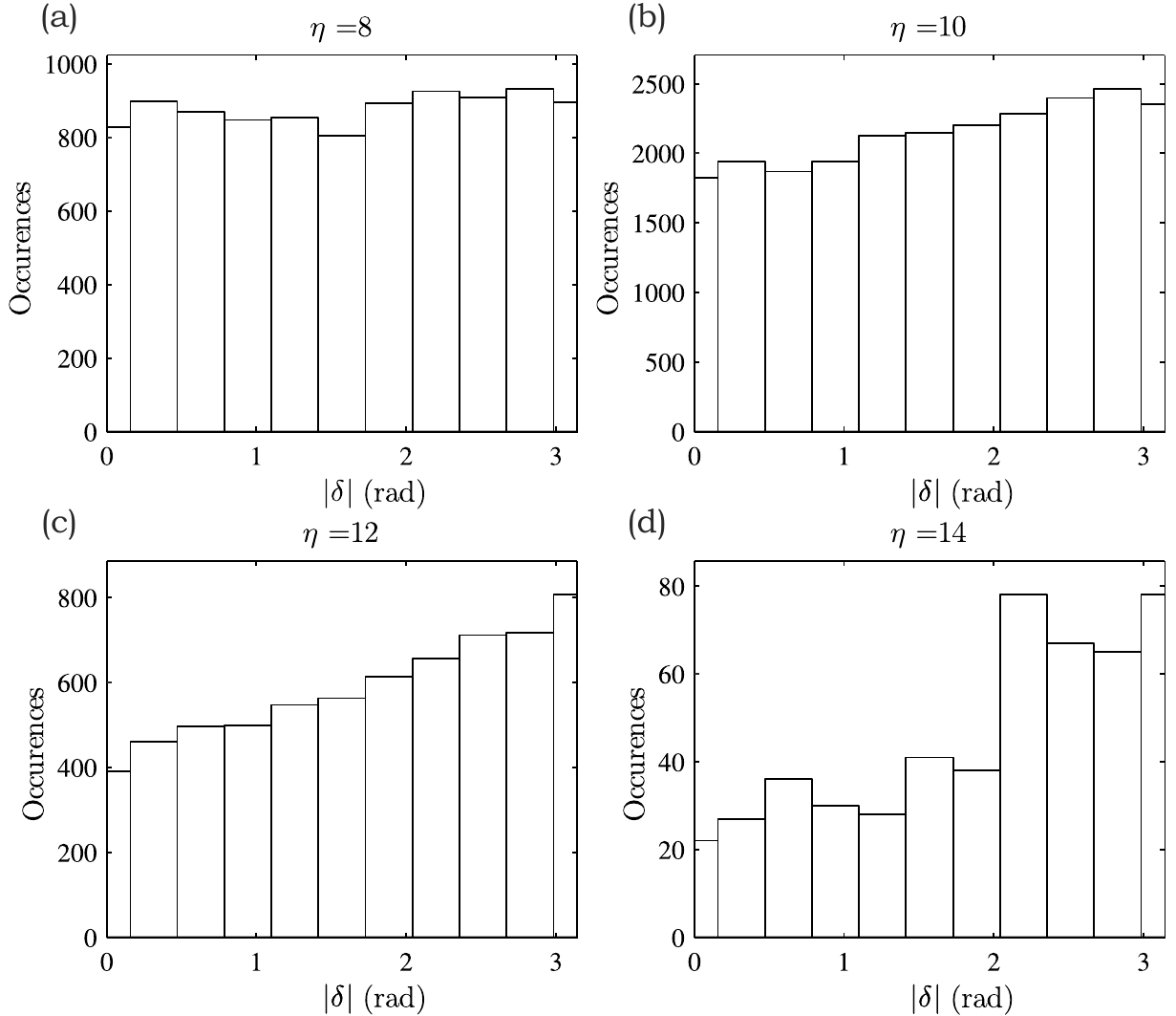}
\caption{\textbf{Simulated phase difference $\abs{\delta}$ for the balanced beam splitter for different intensity enhancements $\eta$ and $\rm{Dim}(\textbf{S})=300$} (bars). The probability that $\delta \rightarrow \pi$ increases with $\eta$. $(a)$ Realizations with $7.5 \leq \eta < 8.5$. $(b)$ Realizations with $9.5 \leq \eta < 10.5$. $(c)$ Realizations with $11.5 \leq \eta < 12.5$. $(d)$ Realizations with $13.5 \leq \eta < 14.5$.} 
\label{Fig_7_9}
\end{figure}

Therefore we have simulated our experiment by applying our optimization algorithm on large random $\textbf{S}$. Fig \ref{Fig_7_9} shows the observed distribution for $\abs{\delta}$ as a function of intensity enhancement $\eta$. The main observation is that $\rm{P}(\abs{\delta})$ has a global maximum at $\abs{\delta}=\pi$ that increases with $\eta$.
The simulations in Fig. \ref{Fig_7_9} are performed for $\rm{Dim}(\textbf{S})=300$. We control the phase of 40 input elements representing incident mode 1, and 40 input elements representing incident mode 2. Each controlled input element has a normalized input power of $1$. We have set margins $\epsilon_1=0$ and $\epsilon_2=0.001$. We apply the optimization algorithm 2500 times per mode to guarantee convergence. We select output elements for which the total power of the optimization for mode 1 is within $10\%$ of the optimization for mode 2, approximating our experiment. The intensity enhancement $\eta$ is given by the observed power in a target speckle spot, divided by $40 \times \avg{\abs{s_{a,b}}^2}=40/\rm{Dim}(\textbf{S})$, where the factor 40 comes from the number of channels that are controlled per incident mode. 

It is beyond the scope of this model to match experimental conditions. In particular the large number of channels is difficult to implement. We have repeated our simulations for several $\rm{Dim}(\textbf{S})$ and several amounts of controlled input channels, always demonstrating a global maximum at a $\abs{\delta}=\pi$ that increases with $\eta$. This demonstrates that the two optimized speckle spots approximate better the behavior of a balanced beam splitter with increasing enhancement, using our optimization algorithm. Based on the model with large unitary matrices, it is likely that this result is independent of the type of optimization algorithm used.

We can also explain our findings by considering an analytic model. Assume that the wavefront shaping process has been completed,  where each enhanced speckle spot has an identical intensity enhancement $\eta$. We are free to define a basis for the complete scattering matrix $\textbf{S}$, which includes the SLM and the scattering material. We write $\textbf{S}$ as:

\begin{equation}
\textbf{S}=\frac{1}{\sqrt{N}}{\left[\begin{array}{ccccc} \sqrt{\eta} & \sqrt{\eta} e^{i \alpha_{\rm{in}}} & s_{1,3} & \cdots & s_{1,N}\\ \sqrt{\eta} e^{i \delta} e^{i \alpha_{\rm{out}}} &  \sqrt{\eta}e^{i \alpha_{\rm{in}}} e^{i \alpha_{\rm{out}}} & & \\ s_{3,1} \\ \vdots \\ s_{N,1} & & & & s_{N,N} \end{array}\right]},
\end{equation}

where the first two elements of the input vector correspond with the field in the input modes and the first two elements of the output vector correspond to the field in the output modes. Therefore the relevant elements in $\textbf{S}$ for the beam splitter are the four top left elements. We have chosen a basis where the phase difference $\delta$ of the beam splitter, as observed in our interference experiment, is included in $s_{2,1}$. The phase difference between the input modes $\alpha_{\rm{in}}$ is included in $s_{1,2}$ and $s_{2,2}$, the phase difference between the enhanced speckle spots $\alpha_{\rm{out}}$ is included in $s_{2,1}$ and $s_{2,2}$. Since $\textbf{S}$ has to be unitary, each column of the matrix should be orthogonal to the other columns. Therefore the innerproduct between the first two columns becomes:

\begin{equation}
\eta(1+\exp{i \delta}) e^{-i\alpha_{\rm{in}}}+ \sum_{i=3}^{i=N} s_{i,1} s^*_{i,2}=0.
\end{equation}

We define $B=\sum_{i=3}^{i=N} s_{i,1} s^*_{i,2}$. For $\eta \ll N$ we assume that all elements $s_{i,j}$ still follows the statistics of the elements of a randomly generated unitary matrix, except for the elements describing the beam splitter. We are dealing with a system where $N \gg 2$ and therefore we can approximate $N-2 \approx N$. We assume that for a random scattering matrix all elements $s_{i,j}$ are complex Gaussian distributed with mean $\avg{s_{i,j}}=0$ and standard deviation $\sigma_{s}=\avg{\abs{s_{i,j}}}=1$. From the rules of multiplication and adding Gaussian distributions it follows that $B$ should be a complex Gaussian distributed value with $\avg{B}=0$ and $\sigma_{B}=\avg{\abs{B}}=\sqrt{\frac{N}{2}}$. Therefore if $\eta > \sqrt{\frac{N}{2}}$, $B$ can be ignored in Eq. (7) and we expect $\delta \rightarrow \pi$ to satisfy this equation. For the simulations in Fig. \ref{Fig_7_9} this should be the case for $\eta=12.2$, which we indeed observe in the simulated distributions. In our experiments we have $\eta \sim 5$ and $N \sim 10^3$, and therefore according to this model $\delta \rightarrow \pi$ for $\eta \sim 10^1$. Therefore our experimental observations of Fig. \ref{Fig_7_7} are not convincingly explained by this model. On the other hand, Eq. (7) sets restrictions on the allowed combinations of $B$, $\alpha_{\rm{in}}$, $\eta$, and $\delta$, which we have ignored up till now. Therefore more advanced modeling is required that is outside the scope of the present paper. 

%
%

\section{Discussion}

We have experimentally created two optimized speckle spots that are correlated like the output of a lossy balanced beam splitter. The interference experiment suggests that $\abs{\delta} \rightarrow \pi$ with increasing $\eta$, which is indicated by a random matrix model. The model could not match the dimension of our experiment, nevertheless, the agreement is gratifying.  Our model demonstrates that the probability distribution for $\delta$ depends on the number of modes in the system. It would be intriguing to perform these kind of experiments with systems of lower dimension, such as, multi-mode fibers with embedded disorder to confirm this scaling \cite{Puente2011}. 

It would be fascinating to measure the transmission matrix of the sample prior to optimization. This allows for an experimental study on the influence of the optimization algorithm and $\eta $ on the observed $\delta$. In addition this would also allow to address speckle patterns with more complicated correlations.

The loss in our experiment can be reduced by several orders of magnitude by implementing more efficient wavefront-shaping procedures and using continuous-wave lasers. On the other hand, our pulsed experiment reveals the opportunity to apply this beam splitter on incident light produced in nonlinear processes, such as entangled photon pairs or higher order quantum states produced with spontaneous parametric down-conversion \cite{Huisman2009, Huisman2013}. This makes it possible to program linear optical circuits for incident quantum states to exploit quantum correlations in disordered media \cite{Lodahl2005, Smolka2009, Bromberg2009, Peruzzo2010, Lahini2010, Ott2010, Peeters2010, Bonneau2012}.

\section{Conclusions and outlook}
In summary, we have controlled light propagation in opaque scattering media with phase modulation of the two incident light modes. We have optimized two speckles that show interference like a 50:50 beam splitter. 
Advanced optimization algorithms make it possible to create more complicated linear circuits out of multiple-scattering media \cite{ThesisHuisman2013}.

\section*{Acknowledgments}
We thank J. P. Korterik, F. G. Segerink, M. P. van Exter, J. L. Herek, I. M. Vellekoop, and  W. L. Vos for discussions and support. A.P.M. acknowledges ERC grant 279248. P.W.H.P. acknowledges NWO Vici.

\end{document}